\documentclass[a4paper]{article}

\usepackage{INTERSPEECH2020}

\usepackage[table,xcdraw]{xcolor}
\usepackage{soul}
\usepackage{url}
\usepackage{subfigure}

\title{CONVOLUTIONAL SPEECH RECOGNITION WITH PITCH \\ AND VOICE QUALITY FEATURES}
\name{Guillermo C\'ambara$^{1,4}$, Jordi Luque$^{2,4}$ and Mireia Farr\'us$^3$}
\address{$^1$Universitat Pompeu Fabra, Barcelona, Spain\\
$^2$Universitat Polit\`ecnica de Catalunya, Barcelona, Spain\\
$^3$Universitat de Barcelona, Barcelona, Spain \\
$^4$Telef\'onica Research, Barcelona, Spain}
\email{guillermo.cambara@upf.edu, jordi.luque@telefonica.com, mfarrus@ub.edu}

\begin{document}
\maketitle
\begin{abstract}
  

The effects of adding pitch and voice quality features such as jitter and shimmer to a state-of-the-art CNN model for Automatic Speech Recognition are studied in this work. Pitch features have been previously used for improving classical HMM and DNN baselines, while jitter and shimmer parameters have proven to be useful for tasks like speaker or emotion recognition. Up to our knowledge, this is the first work combining such pitch and voice quality features with modern convolutional architectures, showing improvements up to 7\% and 3\% relative WER points, for the publicly available Spanish Common Voice and LibriSpeech 100h datasets, respectively. Particularly, our work combines these features with mel-frequency spectral coefficients (MFSCs) to train a convolutional architecture with Gated Linear Units (Conv GLUs). Such models have shown to yield small word error rates, while being very suitable for parallel processing for online streaming recognition use cases. 
We have added pitch and voice quality functionality to Facebook's wav2letter speech recognition framework, and we provide with such code and recipes to the community, to carry on with further experiments. Besides, to the best of our knowledge, our Spanish Common Voice recipe is the first public Spanish recipe for wav2letter.

\end{abstract}

\section{INTRODUCTION}
Neural network models applied to automatic speech recognition (ASR) task are consistently achieving state-of-the-art results in the field. Some of the best scoring architectures involve transformer-based acoustic models \cite{Wang_2020}, LAS models \cite{las2015} with SpecAugment data augmentation \cite{Park_2019} or models strongly based on convolutional neural networks, like the ResNets and TDS ones in \cite{synnaeve2019endtoend}. 

Such convolutional approaches have the advantage of being able to look at larger context windows, without the risk of vanishing gradients like in pure LSTM approaches, and being suitable for online streaming applications, while attaining low word error rate (WER) scores. Furthermore, following the trend of making systems as end-to-end as possible, even fully convolutional neural approaches have been proposed, and shown state-of-the-art performances \cite{zeghidour2018fully}. This fully convolutional architecture takes profit of stacking convolutional layers for efficient parallelization with gated linear units that prevent the gradients from vanishing as architectures go deeper \cite{convGLU}. 

Recently, Facebook has outsourced wav2letter \cite{wav2letterPaper}, a very fast speech recognition framework with recipes prepared for training and decoding with some of these modern models, with an emphasis on the convolutional ones. Most of the modern architectures work only on cepstral (MFCCs) and mel-frequency spectral coefficients (MFSCs) inputs, or even directly with the raw waveform, and tending towards the increasing of granularity at input level by usually augmenting the number of spectral parameters. Whilst it seems evident whether current end-to-end deep network architectures are able to automatically perform relevant feature extraction for speech tasks, psychical or functional properties, related to the underlying speech production system, become fuzzy or difficult to connect with the speech recognition performances. In addition, it is still unclear how the great quantity and different speech hand-crafted voice features, carefully developed along past years and based on our linguistic knowledge, might help and in which degree to the current speech network architectures. 

Some well-known speech recognition frameworks, like Kaldi \cite{kaldi}, have incorporated the use of additional prosodic features, such as the pitch or the probability of voicing (POV). These are stacked into an input vector together with those cepstral/spectral ones, and then forwarded to classifiers like HMM or DNN ensembles. Nevertheless, the newest convolutional architectures have not yet been extensively applied along with such prosodic features, and frameworks like wav2letter, reaching state-of-the-art performances in ASR tasks, do not yet provide with integrated pitch functionality within feature extraction modules. Furthermore, in the last decades, jitter and shimmer have shown to be useful in a wide range of applications; e.g. detection of different speaking styles \cite{slyh1999analysis}, age and gender classification \cite{wittig2003implicit}, emotion detection \cite{li2007stress}, speaker recognition \cite{farrus2007jitter}, speaker diarization \cite{zewoudie2014jitter}, and Alzheimer's and Parkinson Diseases detection \cite{mirzaei2018two,benba2014hybridization}, among others. 

The main contribution of this work focus on assessing the value of adding pitch and voice quality features to the spectral coefficients, commonly employed in most of the deep speech recognition systems. To this end, the dimension of MFSC vector, at the input layer, is augmented by the prosodic features and error rates are reported for both Spanish and English speech recognition tasks. Experiments are carried out by using the Conv GLU model. It was proposed by \cite{wav2letterWSJRecipe} within the wav2letter's WSJ recipe and has reported state-of-the-art performances for both LibriSpeech and WSJ datasets. To the best of our knowledge, this is the first attempt to use jitter and shimmer features within a modern deep neural-based speech recognition system while keeping {\it easy to identify} psychical/functional properties of the voice and linking them to the ASR performance. Furthermore, the recipes employed in this research have been published in a GitHub\footnote{\url{https://github.com/gcambara/wav2letter/tree/wav2letter_pitch}} repository, aiming to make easy to reproduce experiments on two public and freely available corpus: the Spanish Common Voice dataset \cite{ardila2019common} and the English LibriSpeech 100h partition \cite{librispeech}.



\section{PROSODIC AND VOICE QUALITY FEATURES}\label{voice}


Pitch --or fundamental perceived frequency-- has proved to increase performance in ASR systems, significantly for tonal languages, like Punjabi \cite{GUGLANI2020107386}, as well as for non-tonal ones like English \cite{pitchASR}. Jitter and shimmer represent the cycle-to-cycle variations of the fundamental frequency and amplitude, respectively. Since a long time, they have been considered relevant and often applied to detect voice pathologies, as in \cite{kreiman2005perception,michaelis1998some}, thus considered as measurements of voice quality. Although voice quality features differ intrinsically from those suprasegmental prosodic features, they have shown to be related to prosody. In \cite{campbellvoice}, the authors showed that voice quality features are relevant markers signaling paralinguistic information, and that they should even be considered as prosodic carriers along with pitch and duration, for instance. 

In ASR literature, some works have reported that prosodic information can raise the performance of speech recognizer systems. For instance, in \cite{zaidi2019automatic} the authors built an ASR for dysarthric speech and \cite{robustJitter} reports benefit on the use of jitter and shimmer for noisy speech recognition, both systems based on classical HMM acoustic modelling. In the case of deep networks,  LSTMs was taken into \cite{cho2019deep} for acoustic emotion recognition task, however they did not perform ASR task on its own.

Following previous works, we hypothesize that prosodic and voice quality features may boost robustness in convolutional like ASR systems. Moreover, they could play an even more important role in further speech tasks, including punctuation marks, emotion recognition or musical contexts, where additional prosodic information would be useful.

\section{METHODOLOGY}\label{experiments}

\subsection{Data}

The effect of adding pitch and voice quality features is evaluated by means of the Common Voice corpus in Spanish \cite{ardila2019common} and the LibriSpeech 100h dataset in English \cite{librispeech}. Common Voice corpus is an open-source dataset that consists of recordings from volunteer contributors pronouncing scripted sentences, recorded at 48kHz rate and using own devices. The sentences come from original contributor donations and public domain movie scripts and it is continuously growing. 
Although there are already more than 100 hours of validated audio, we have kept a reduced partition of approximately 19.0 h for training, 2.7 h for development and 2.2 h for testing sets. The main criterion for the stratification of such partitions is to ensure that each one has exclusive speakers, while trying to keep a 80-10-10\% proportion. Every sample can be down voted by the contributors if it is not clear enough, so we have discarded all samples containing at least one down vote, to keep the cherry picked recordings as clean as possible. Afterwards, we try to keep as balanced as possible the distributions by age, gender and accent. The Python scripts for obtaining such partition are provided in our public Git repository, along with other code necessary to reproduce our ASR recipes. Up to our knowledge, this is the first public repository with a wav2letter recipe for a publicly available Spanish dataset. Besides, in order to provide with results for a popular benchmark, the proposal is also assessed with the aforementioned LibriSpeech 100h partition in English, consisting of audio book recordings sampled at 16kHz.

\subsection{Feature Extraction}

As recommended by wav2letter's Conv GLU recipes, raw audio is processed to extract static MFSCs, applying 40 filterbanks. This serves as our baseline, so on top of it we append pitch and voice quality related features. From now on in this work, when we talk about pitch features we refer to the following three features: the extracted pitch itself, plus the POV for each frame and the variation of pitch across two frames (delta-pitch). Being so, 40 MFSCs are always computed for each time frame, and if specified by the user in the configuration, the three pitch features (pitch, POV and delta-pitch) can be appended to them, plus jitter relative, jitter absolute, shimmer dB and/or shimmer relative.

There are various pitch extractor algorithms such as Yin \cite{yinPitch} or getF0 \cite{Talkin2005ARA}. However, we have decided to refactor the Kaldi's one from \cite{Ghahremani2014APE} within the feature extractor C++ class from wav2letter. Latter algorithm has been frequently tested along the recent years within a wide variety of ASR tasks. 
It is inspired by getF0 and finds the sequence of lags that maximizes the Normalized Cross Correlation Function (NCCF). It makes use of the Viterbi algorithm for obtaining the optimal lags and, in our implementation, it applies the logarithm to the pitch values as the only post-processing step. The logarithm compresses pitch values to the same order as the MFSCs, which are compressed by the logarithm as well, thus improving numerical stability later on during the training phase. Subtracting the weighted average pitch during post-processing has been discarded, since the reported gains in WER by Kaldi are only of a 0.1\%, but we may implement them in future iterations.

Shimmer is computed measuring the peak-to-peak waveform amplitude at each period where the pitch is extracted, and then performing the corresponding operations, depending on whether we deal with shimmer dB or shimmer relative, see reference \cite{farrus2007jitter}. With the pitch extracted at each period, the same can be done for jitter absolute and relative, by calculating the fundamental frequency differences between such cycles. 

\subsection{System Architecture}

Since our purpose is to study how pitch and voice quality features contribute to a convolutional acoustic model (AM), we have used the Conv GLU AM from wav2letter's Wall Street Journal (WSJ) recipe \cite{wav2letterWSJRecipe}. This model has approximately 17M parameters with dropout applied after each of its 17 layers. The WSJ dataset contains around 80 hours of audio recordings, which is closer to the magnitude of our data than the full LibriSpeech recipe (about 1000 hours). We have not done an extensive exploration of architecture parameters, since it yields decent out of the box results with Common Voice and LibriSpeech 100h data.

\begin{table*}[]
\renewcommand{\arraystretch}{1.50}
\caption{WER percentages by augmenting spectral features with prosody and voice quality ones. The results are reported on the Common Voice's development (Dev) and test (Test) sets, comprising 2.7 hours and 2.2 hours, respectively. Error rates are obtained by using a greedy decoding without language model (NoLM) and by a beam search decoding using a 4-gram LM trained both on the Common Voice's training subset (CVLM) and on the training partition of the Spanish corpus Fisher-Callhome (FCLM).}
\vspace{-0.5cm}
\label{commonvoice_wer_table}
\small{
\begin{center}
\begin{tabular}{lcccccc}
\hline
  \multicolumn{1}{c}{AM}  & \multicolumn{6}{c}{WER (\%)}   \\
\cline{1-1} \cline{2-7}
\bfseries Features & \bfseries NoLM-Dev & \bfseries NoLM-Test & \bfseries CVLM-Dev & \bfseries CVLM-Test & \bfseries FCLM-Dev & \bfseries FCLM-Test\\
\hline\hline
MFSC & $64.92$ & $70.07$ & $20.29$ & $24.72$ & $38.58$ & $44.20$ \\
\hspace{5mm}  + Pitch & \boldsymbol{$63.18$} & \boldsymbol{$68.79$} &  $20.56$ & $24.89$ & \boldsymbol{$37.57$} & $43.18$ \\
\hspace{5mm} + Pitch + Jitter & $63.83$ & $69.56$ &  $20.28$ & $23.97$ & $38.07$ & $43.26$ \\
\hspace{5mm} + Pitch + Shimmer & $73.18$ & $77.04$ &  $23.30$ & $25.10$ & $46.90$ & $50.60$ \\
\hspace{5mm} + Pitch + Jitter + Shimmer & $64.46$ & $69.51$ & \boldsymbol{$20.01$} & \boldsymbol{$22.90$} & $38.63$ & \boldsymbol{$42.95$}\\

\hline
\end{tabular}
\vspace{-0.6cm}
\end{center}
}
\end{table*}

Regarding Common Voice's lexicon, we use a grapheme-based one extracted from the approximately 9000 words from both the training and development partitions. We use the standard Spanish alphabet as tokens, plus the "ç" letter from Catalan and the vowels with diacritical marks, making a total of 37 tokens. The "ç" character is included because of the presence of some Catalan words in the dataset, like "Barça". The language model (LM) is a 4-gram model extracted with KenLM \cite{kenLM} from the training set. Since most of the sentences are shared across partitions, due to the scripted nature of the dataset, we expected an optimistic behavior after applying such LM. Therefore, we are also reporting results given by another 4-gram LM extracted from the Spanish Fisher+Callhome. The Fisher corpus splitting is taken from the Kaldi's recipe \cite{weiss2017sequencetosequence}. Decoding across AM, lexicon and LM is done with the beam-search decoder provided by wav2letter \cite{Liptchinsky2017LetterBasedSR}. 
Furthermore, in order to assess the capacity of the AM by itself, we also evaluate without LM, choosing the final characters with the greedy best path from the predictions of the AM.
For the LibriSpeech evaluation, the lexicon and the language model are the same as provided by wav2letter's Conv GLU LibriSpeech recipe. The lexicon is obtained from the train corpus and the language model is a 4-gram model also trained with KenLM.

\subsection{Experiments}

After some initial simulations, we have found that the most stable voice quality features are jitter relative and shimmer relative. Therefore, we try 5 different feature configurations:

\begin{enumerate}
\item 40 MFSCs only
\vspace{-0.4em}
\item 40 MFSCs + 3 pitch features
\vspace{-0.4em}
\item 40 MFSCs + 3 pitch + 1 relative jitter
\vspace{-0.4em}
\item 40 MFSCs + 3 pitch + 1 rel. shimmer
\vspace{-0.4em}
\item 40 MFSCs + 3 pitch + 1 rel. jitter + 1 rel. shimmer
\end{enumerate}

For each one, we compute WERs on Common Voice's dev and test sets. Decodings are performed without LM (NoLM), with both in-domain and out-domain LMs, from Common Voice's LM (CVLM) from Fisher+Callhome's LM (FCLM) databases, respectively. Therefore, we obtain 6 WERs for each one of the 5 feature configurations.

Besides the features, the training configurations for each experiment are the same, all based on wav2letter's WSJ recipe. The inferred segmentation is taken out from wav2letter's Auto Segmentation Criterion (ASG) \cite{wav2letterWSJRecipe}, inspired by CTC loss \cite{ctc}. The learning rate is tweaked to $7.3$, and is decayed in a 20\% every 10 epochs, a tuning done with the dev set. A 25 ms rolling window with a 10 ms stride is used for extracting all the features, jitter and shimmer are averaged across 500 ms windows. 

For beam-search decoding, the following settings are tuned with the dev set: LM weight set to 2.5, word score set to 1, beam size set to 2500, beam threshold set to 25 and silence weight set to -0.4. In order to tune these, we have not run an extensive exploration of hyperparameters, but after a shallow search we found these to provide good results for both LMs.

Furthermore, LibriSpeech WER is evaluated with dev-clean/other and test-clean/other partitions, using the same AM training recipe as in Common Voice. As we are scaling with a bigger dataset demanding a higher computational cost, the top three parameter configurations found with Common Voice experiments are selected in order to perform such evaluations. Decoding parameters are taken from wav2letter's LibriSpeech recipe.

\vspace{-0.4cm}
\section{RESULTS AND DISCUSSION}\label{results}

The Table \ref{commonvoice_wer_table} reports the word error rates (WER, \%) for each one of the 5 feature configurations, for the proposed decodings of Common Voice's dev and test sets, without LM (NoLM), with its own LM (CVLM) and the Fisher+Callhome LM (FCLM). For every evaluated case, the best WER score is always provided by one of the models using pitch features, or pitch with voice quality (jitter + shimmer) features, with gains between 1.38\% and 7.36\% relative WER points.

For the cases without LM, the model with MFSC and pitch features is the one with the best performance, with relative gains of 2.68\% and 1.83\% for dev and test sets, respectively. Additional features on the other models also improve the WER score, except for the case with pitch and shimmer only, which yields worse results across all experiments. On the other hand, decoding with CVLM achieves the best WER scores, when training with all the proposed features together: MFSCs, the 3 pitch features, jitter relative and shimmer relative. A 20.01\% WER is obtained for the dev set, and a 22.90\% WER for the test set. As it was expected, the CVLM improves drastically the predictions, because even though it is obtained from the train partition solely, many sentences are shared with the dev and test sets, due to the reduced vocabulary in this dataset.

A more realistic approach is to decode by using an external LM. The FCLM language model is built from the training partition of the LDC Spanish Fisher+Callhome corpus. Although the LM enrollment is performed with less than 20 hours of audio (approximately 16k sentences), it still yields to a reasonable performance compared to the CVLMs decodings. With respect to the prosodic features, the FCLM beam decoding reaches the lower WER in development by using MFSCs only augmented with pitch features; that is, 37.57\% WER. The lowest 42.95\% WER score in the test set is given by the combination of all pitch and voice quality characteristics. Once again, the best results in terms of WER are provided by models with pitch features, or pitch features with the combination of jitter and shimmer, showing the potential of pitch and voice quality features to improve the performance of an ASR based on convolutional neural networks. 

\begin{figure}[t!]
\centering     
\hspace{-0.3cm}
\begin{subfigure}{\label{fig:cv_asr_training}\includegraphics[width=0.52\linewidth]{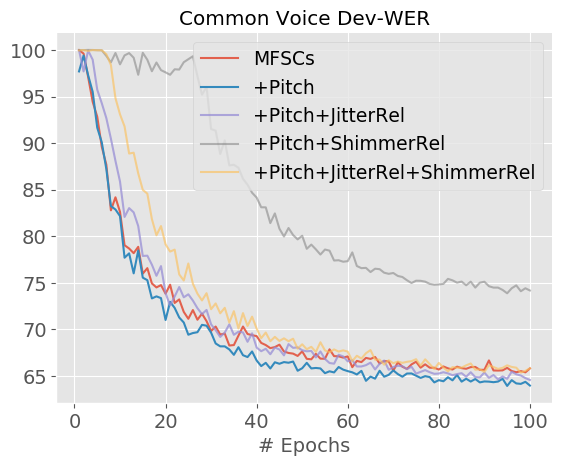}}
\hspace{-0.35cm}
\end{subfigure}
\subfigure{\label{fig:libri_asr_training}\includegraphics[width=0.52\linewidth]{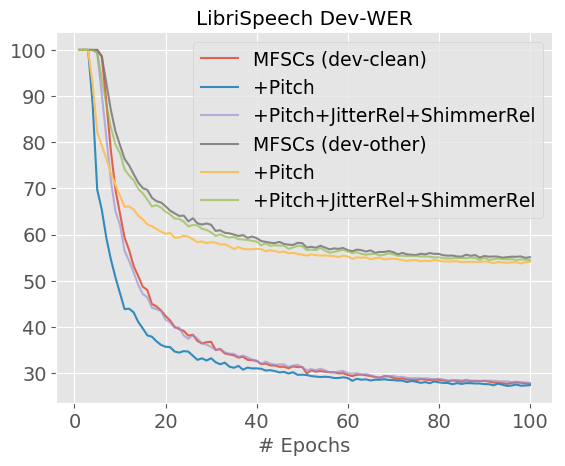}}
\vspace{-0.3cm}
\caption{Common Voice and LibriSpeech dev set WER (\%) during training, as a function of the epoch number. For the latter, dev-clean and dev-other are evaluated. Curves across the 5 different feature configurations for the same acoustic model architecture.}
\vspace{-0.3cm}
\end{figure}

Nonetheless, it is worth noticing how the use of only pitch and shimmer features yields to worse performance for both AM and AM/LM decoding models. Previous behaviour is depicted in the Figure \ref{fig:cv_asr_training}, where using only shimmer dramatically affects the training stage of the model, making it worse and slower. However, training with pitch features or with pitch and jitter features seems to help at reaching better WER plateaus and at faster pace. While jitter is a measure of frequency instability in the wave, shimmer is a measure of amplitude instability. Being so, pitch and jitter characteristics might contribute to MFSCs spectral features with independent information, just by synchronising them in a simple concatenation like the proposed one. However, the inclusion of shimmer, which is related to amplitude --as opposed to the pitch and jitter, related to frequency--, is more likely to be understood as a perturbation throughout the convolutional layers that might difficult the acoustic model training. 

Even though, it is interesting to see how if shimmer is coupled with jitter and pitch characteristics altogether, the performance obtained yields to more robust results compared to the baseline and independently of decoding with CVLM and FCLM language models. Other studies already suggest the correlation between jitter and shimmer by the same index, that is, the Voice Handicap Index (VHI) \cite{jitterShimmerCorrelation}, so the convolutional filters may be finding similar correlations, thus improving mutual information when coupled together with spectral features and promoting such as voice measurements as good feature candidates for enhancing the speech recognition of pathological voices. The latter being an interesting hypothesis to look for further evidence.

The impact of pitch and voice quality features is also reported by LibriSpeech experiments, see Table \ref{librispeech_wer_table} and Figure \ref{fig:libri_asr_training}, with relative improvements of 2.94\% and 2.06\% for dev-clean and dev-other, respectively, and about 0.96\% and 2.87\% for test-clean and test-other. Gains seem to be more consistent for the "other" partitions, where there is more accent and prosody diversity than in "clean" ones.

Appending pitch characteristics to MFSCs seems to slightly improve the ASR performance. Among them, MFSC+pitch and MFSC+pitch+jitter+shimmer combinations are the ones that provide the most robust behavior across all the experiment. All assessed features carry prosodic information and might aid the network on complementing the information conveyed by solely the magnitude spectrum. For instance, by helping on reducing the MFSC distortion which appears at the lower frequency region of the spectrum~\cite{yadav2019}. Overall, they help boosting the performance of the convolutional acoustic model for both the different databases and the languages studied in this work. 


Note that the approach employed for the feature combination has been quite simple, by just appending such features to the spectral ones at the input layer, without extensive post-processing either nor adaptation of the model architecture. Being so, it is reasonable to think that there is still margin of improvement in the application of pitch and voice quality measurements to state-of-the-art convolutional neural models. Possible strategies comprises adapting the feature concatenation, maybe by dedicating exclusive filters to the new pitch and voice quality features with POV as a gating mechanism, especially after experimentally realising, not reported in this work, that the estimation of measurements like shimmer may benefit from different post-processing techniques.

\begin{table}[t]
\setlength{\tabcolsep}{3pt}
\renewcommand{\arraystretch}{1.50}
\caption{LibriSpeech WER values for the three best performing combinations of features proposed in the Acoustic Model (AM) in the Common Voice experiments: MFSC, MFSC+Pitch and MFSC+Pitch+Shimmer+Jitter (shortened as "All"). Decoding is done with a 4-gram LM trained with LibriSpeech train set transcripts.}
\vspace{-0.5cm}
\label{librispeech_wer_table}
\small{
\begin{center}
\begin{tabular}{lcccc}
\hline
  \multicolumn{1}{c}{AM}  & \multicolumn{4}{c}{WER (\%)}   \\
\cline{1-1} \cline{2-5}
\bfseries Features & \bfseries dev-clean & \bfseries dev-other & \bfseries test-clean & \bfseries test-other\\
\hline\hline
MFSC & $10.22$ & $31.59$ & $10.38$ & $34.46$ \\
\hspace{5mm}+ Pitch & $9.94$ & $30.95$ & \boldsymbol{$10.28$} & \boldsymbol{$33.47$} \\
\hspace{5mm} + All & \boldsymbol{$9.92$} & \boldsymbol{$30.94$} & $10.37$ & $33.57$ \\
\hline
\end{tabular}
\vspace{-0.6cm}
\end{center}
}
\end{table}
\vspace{-0.2cm}
\section{CONCLUSIONS}\label{conclusions}

In this study, we performed a preliminary exploration on the effects of pitch and jitter/shimmer voice quality measurements within the framework of the ASR task performed by convolutional neural network models. The experiments reported with a publicly available Spanish speech corpus showed consistent improvements on the model robustness, achieving a reduced relative 7\% WER in some scenarios. Besides, such feature extraction functionalities are provided and integrated with wav2letter code to easily replicate our findings or directly applying pitch and voice quality features to wav2letter models. We also provide the recipe for the Common Voice Spanish dataset, the first recipe suited for wav2letter using a Spanish publicly available dataset. The recipe for LibriSpeech experiments is also provided, which achieves up to a 2.94\% relative WER improvement. Further steps on the research of convolutional ASR with pitch and voice quality would imply adapting architectures for feature processing, or applying such characteristics for tasks including the presence of punctuation marks, emotion recognition and even pathological or singing voices.  
For the latter tasks, the importance of pitch and voice quality features is expected to become even more relevant.

\vspace{-0.2cm}
\section{ACKNOWLEDGMENTS}

This work is a part of the INGENIOUS project, funded by the European Union’s Horizon 2020 Research and Innovation Programme and the Korean Government under Grant Agreement No 833435. The third author has been funded by the Agencia Estatal de Investigación (AEI), Ministerio de Ciencia, Innovación y Universidades and the Fondo Social Europeo (FSE) under grant RYC-2015-17239 (AEI/FSE, UE).

\bibliographystyle{IEEEtran}

\bibliography{mybib}

\begin{thebibliography}{10}
\providecommand{\url}[1]{#1}
\csname url@samestyle\endcsname
\providecommand{\newblock}{\relax}
\providecommand{\bibinfo}[2]{#2}
\providecommand{\BIBentrySTDinterwordspacing}{\spaceskip=0pt\relax}
\providecommand{\BIBentryALTinterwordstretchfactor}{4}
\providecommand{\BIBentryALTinterwordspacing}{\spaceskip=\fontdimen2\font plus
\BIBentryALTinterwordstretchfactor\fontdimen3\font minus
  \fontdimen4\font\relax}
\providecommand{\BIBforeignlanguage}[2]{{%
\expandafter\ifx\csname l@#1\endcsname\relax
\typeout{** WARNING: IEEEtran.bst: No hyphenation pattern has been}%
\typeout{** loaded for the language `#1'. Using the pattern for}%
\typeout{** the default language instead.}%
\else
\language=\csname l@#1\endcsname
\fi
#2}}
\providecommand{\BIBdecl}{\relax}
\BIBdecl

\bibitem{Wang_2020}
Y.~Wang, A.~Mohamed, D.~Le, C.~Liu, A.~Xiao, J.~Mahadeokar, H.~Huang,
  A.~Tjandra, X.~Zhang, F.~Zhang \emph{et~al.}, ``Transformer-based acoustic
  modeling for hybrid speech recognition,'' in \emph{ICASSP 2020-2020 IEEE
  International Conference on Acoustics, Speech and Signal Processing
  (ICASSP)}.\hskip 1em plus 0.5em minus 0.4em\relax IEEE, 2020, pp. 6874--6878.

\bibitem{las2015}
\BIBentryALTinterwordspacing
W.~Chan, N.~Jaitly, Q.~V. Le, and O.~Vinyals, ``Listen, attend and spell,''
  \emph{CoRR}, vol. abs/1508.01211, 2015. [Online]. Available:
  \url{http://arxiv.org/abs/1508.01211}
\BIBentrySTDinterwordspacing

\bibitem{Park_2019}
D.~S. Park, W.~Chan, Y.~Zhang, C.-C. Chiu, B.~Zoph, E.~D. Cubuk, and Q.~V. Le,
  ``Specaugment: A simple data augmentation method for automatic speech
  recognition,'' \emph{arXiv preprint arXiv:1904.08779}, 2019.

\bibitem{synnaeve2019endtoend}
G.~Synnaeve, Q.~Xu, J.~Kahn, T.~Likhomanenko, E.~Grave, V.~Pratap, A.~Sriram,
  V.~Liptchinsky, and R.~Collobert, ``End-to-end asr: from supervised to
  semi-supervised learning with modern architectures,'' 2019.

\bibitem{zeghidour2018fully}
N.~Zeghidour, Q.~Xu, V.~Liptchinsky, N.~Usunier, G.~Synnaeve, and R.~Collobert,
  ``Fully convolutional speech recognition,'' 2018.

\bibitem{convGLU}
\BIBentryALTinterwordspacing
Y.~N. Dauphin, A.~Fan, M.~Auli, and D.~Grangier, ``Language modeling with gated
  convolutional networks,'' \emph{CoRR}, vol. abs/1612.08083, 2016. [Online].
  Available: \url{http://arxiv.org/abs/1612.08083}
\BIBentrySTDinterwordspacing

\bibitem{wav2letterPaper}
V.~Pratap, A.~Hannun, Q.~Xu, J.~Cai, J.~Kahn, G.~Synnaeve, V.~Liptchinsky, and
  R.~Collobert, ``wav2letter++: The fastest open-source speech recognition
  system,'' 12 2018.

\bibitem{kaldi}
D.~Povey, A.~Ghoshal, G.~Boulianne, L.~Burget, O.~Glembek, N.~Goel,
  M.~Hannemann, P.~Motlíček, Y.~Qian, P.~Schwarz, J.~Silovský, G.~Stemmer,
  and K.~Vesel, ``The kaldi speech recognition toolkit,'' \emph{IEEE 2011
  Workshop on Automatic Speech Recognition and Understanding}, 01 2011.

\bibitem{slyh1999analysis}
R.~E. Slyh, W.~T. Nelson, and E.~G. Hansen, ``Analysis of mrate, shimmer,
  jitter, and f/sub 0/contour features across stress and speaking style in the
  susas database,'' in \emph{1999 IEEE International Conference on Acoustics,
  Speech, and Signal Processing. Proceedings. ICASSP99 (Cat. No. 99CH36258)},
  vol.~4.\hskip 1em plus 0.5em minus 0.4em\relax IEEE, 1999, pp. 2091--2094.

\bibitem{wittig2003implicit}
F.~Wittig and C.~M{\"u}ller, ``Implicit feedback for user-adaptive systems by
  analyzing the users' speech,'' in \emph{Proceedings of the Workshop on
  Adaptivität und Benutzermodellierung in interaktiven Softwaresystemen
  (ABIS)}, Karlsruhe, Germany, 2003.

\bibitem{li2007stress}
X.~Li, J.~Tao, M.~T. Johnson, J.~Soltis, A.~Savage, K.~M. Leong, and J.~D.
  Newman, ``Stress and emotion classification using jitter and shimmer
  features,'' in \emph{2007 IEEE International Conference on Acoustics, Speech
  and Signal Processing-ICASSP'07}, vol.~4.\hskip 1em plus 0.5em minus
  0.4em\relax IEEE, 2007, pp. IV--1081.

\bibitem{farrus2007jitter}
M.~Farr{\'u}s, J.~Hernando, and P.~Ejarque, ``Jitter and shimmer measurements
  for speaker recognition,'' in \emph{Proceedings of the Interspeech}, Antwerp,
  Belgium, 2007.

\bibitem{zewoudie2014jitter}
A.~W. Zewoudie, J.~Luque, and F.~J. Hernando~Peric{\'a}s, ``Jitter and shimmer
  measurements for speaker diarization,'' in \emph{VII Jornadas en
  Tecnolog{\'\i}a del Habla and III Iberian SLTech Workshop: proceedings:
  November 19-21, 2014: Escuela de Ingenier{\'\i}a en Telecomunicaci{\'o}n y
  Electr{\'o}nica Universidad de Las Palmas de Gran Canaria: Las Palmas de Gran
  Canaria, Spain}, 2014, pp. 21--30.

\bibitem{mirzaei2018two}
S.~Mirzaei, M.~El~Yacoubi, S.~Garcia-Salicetti, J.~Boudy, C.~Kahindo,
  V.~Cristancho-Lacroix, H.~Kerherv{\'e}, and A.-S. Rigaud, ``Two-stage feature
  selection of voice parameters for early alzheimer's disease prediction,''
  \emph{IRBM}, vol.~39, no.~6, pp. 430--435, 2018.

\bibitem{benba2014hybridization}
A.~Benba, A.~Jilbab, and A.~Hammouch, ``Hybridization of best acoustic cues for
  detecting persons with parkinson's disease,'' in \emph{2014 Second World
  Conference on Complex Systems (WCCS)}.\hskip 1em plus 0.5em minus 0.4em\relax
  IEEE, 2014, pp. 622--625.

\bibitem{wav2letterWSJRecipe}
\BIBentryALTinterwordspacing
R.~Collobert, C.~Puhrsch, and G.~Synnaeve, ``Wav2letter: an end-to-end
  convnet-based speech recognition system,'' \emph{CoRR}, vol. abs/1609.03193,
  2016. [Online]. Available: \url{http://arxiv.org/abs/1609.03193}
\BIBentrySTDinterwordspacing

\bibitem{ardila2019common}
R.~Ardila, M.~Branson, K.~Davis, M.~Henretty, M.~Kohler, J.~Meyer, R.~Morais,
  L.~Saunders, F.~M. Tyers, and G.~Weber, ``Common voice: A
  massively-multilingual speech corpus,'' 2019.

\bibitem{librispeech}
V.~Panayotov, G.~Chen, D.~Povey, and S.~Khudanpur, ``Librispeech: An asr corpus
  based on public domain audio books,'' 04 2015, pp. 5206--5210.

\bibitem{GUGLANI2020107386}
J.~Guglani and A.~Mishra, ``Automatic speech recognition system with pitch
  dependent features for punjabi language on kaldi toolkit,'' \emph{Applied
  Acoustics}, vol. 167, p. 107386, 2020.

\bibitem{pitchASR}
M.~Magimai-Doss, T.~Stephenson, and H.~Bourlard, ``Using pitch frequency
  information in speech recognition,'' 01 2003.

\bibitem{kreiman2005perception}
J.~Kreiman and B.~R. Gerratt, ``Perception of aperiodicity in pathological
  voice,'' \emph{The Journal of the Acoustical Society of America}, vol. 117,
  no.~4, pp. 2201--2211, 2005.

\bibitem{michaelis1998some}
D.~Michaelis, M.~Fr{\"o}hlich, H.~W. Strube, E.~Kruse, B.~Story, and I.~R.
  Titze, ``Some simulations concerning jitter and shimmer measurement,'' in
  \emph{3rd International Workshop on Advances in Quantitative Laryngoscopy,
  Aachen, Germany}, 1998, pp. 744--754.

\bibitem{campbellvoice}
N.~Campbell and P.~Mokhtari, ``Voice quality: the 4th prosodic dimension,'' pp.
  2417--2420, 2003.

\bibitem{zaidi2019automatic}
B.-F. Zaidi, M.~Boudraa, S.-A. Selouani, D.~Addou, and M.~S. Yakoub,
  ``Automatic recognition system for dysarthric speech based on mfcc’s,
  pncc’s, jitter and shimmer coefficients,'' in \emph{Science and Information
  Conference}.\hskip 1em plus 0.5em minus 0.4em\relax Springer, 2019, pp.
  500--510.

\bibitem{robustJitter}
H.~Rahali, Z.~Hajaiej, and N.~Ellouze, ``Robust features for noisy speech
  recognition using jitter and shimmer,'' \emph{International Journal of
  Innovative Computing, Information and Control}, vol.~11, pp. 955--963, 01
  2015.

\bibitem{cho2019deep}
J.~Cho, R.~Pappagari, P.~Kulkarni, J.~Villalba, Y.~Carmiel, and N.~Dehak,
  ``Deep neural networks for emotion recognition combining audio and
  transcripts,'' 2019.

\bibitem{yinPitch}
A.~de~Cheveigné and H.~Kawahara, ``Yin, a fundamental frequency estimator for
  speech and music,'' \emph{The Journal of the Acoustical Society of America},
  vol. 111, no.~4, pp. 1917--1930, 2002.

\bibitem{Talkin2005ARA}
D.~Talkin, ``A robust algorithm for pitch tracking ( rapt ),'' 2005.

\bibitem{Ghahremani2014APE}
P.~Ghahremani, B.~BabaAli, D.~Povey, K.~Riedhammer, J.~Trmal, and S.~Khudanpur,
  ``A pitch extraction algorithm tuned for automatic speech recognition,''
  \emph{2014 IEEE International Conference on Acoustics, Speech and Signal
  Processing (ICASSP)}, pp. 2494--2498, 2014.

\bibitem{kenLM}
K.~Heafield, ``Kenlm: Faster and smaller language model queries,'' in
  \emph{Proceedings of the Sixth Workshop on Statistical Machine Translation},
  ser. WMT ’11.\hskip 1em plus 0.5em minus 0.4em\relax USA: Association for
  Computational Linguistics, 2011, p. 187–197.

\bibitem{weiss2017sequencetosequence}
R.~J. Weiss, J.~Chorowski, N.~Jaitly, Y.~Wu, and Z.~Chen,
  ``Sequence-to-sequence models can directly translate foreign speech,'' 2017.

\bibitem{Liptchinsky2017LetterBasedSR}
V.~Liptchinsky, G.~Synnaeve, and R.~Collobert, ``Letter-based speech
  recognition with gated convnets,'' \emph{ArXiv}, vol. abs/1712.09444, 2017.

\bibitem{ctc}
A.~Graves, S.~Fern\'{a}ndez, F.~Gomez, and J.~Schmidhuber, ``Connectionist
  temporal classification: Labelling unsegmented sequence data with recurrent
  neural networks.''\hskip 1em plus 0.5em minus 0.4em\relax New York, NY, USA:
  Association for Computing Machinery, 2006.

\bibitem{jitterShimmerCorrelation}
A.~Schindler, F.~Mozzanica, M.~Vedrody, P.~Maruzzi, and F.~Ottaviani,
  ``Correlation between the voice handicap index and voice measurements in four
  groups of patients with dysphonia,'' \emph{Otolaryngology–Head and Neck
  Surgery}, vol. 141, no.~6, pp. 762--769, 2009.

\bibitem{yadav2019}
\BIBentryALTinterwordspacing
I.~C. Yadav, S.~Shahnawazuddin, and G.~Pradhan, ``Addressing noise and pitch
  sensitivity of speech recognition system through variational mode
  decomposition based spectral smoothing,'' \emph{Digit. Signal Process.},
  vol.~86, pp. 55--64, 2019. [Online]. Available:
  \url{https://doi.org/10.1016/j.dsp.2018.12.013}
\BIBentrySTDinterwordspacing

\end{thebibliography}


\end{document}